\documentclass[%
10pt,
 amsmath,amssymb,
 longbibliography,
 aps,
 prd, 
 nofootinbib,
]{revtex4-2}

\usepackage{graphicx}
\usepackage{dcolumn}
\usepackage{bm}
\usepackage{caption}
\usepackage{subcaption} 

\usepackage{amsmath}
\usepackage{hyperref}

\usepackage[toc,page]{appendix}
\usepackage{lipsum}
\usepackage{threeparttable}
\usepackage[bottom]{footmisc}
\usepackage{cleveref}

\begin{document}


\title{The optimal redshift for dark energy I: formalism and interpretation}

\author{Mustapha Ishak}
\email{mishak@utdallas.edu}
\author{Travis Seth Rippentrop}
\email{tsr200000@utdallas.edu}
\author{Kristian Gonzalez}
\email{keg180005@utdallas.edu}

\affiliation{Department of Physics, The University of Texas at Dallas, Richardson, TX 75080, USA}

\date{\today}

\begin{abstract}
We introduce the concept of the optimal redshift for dark energy, a statistically motivated redshift at which departures of the dark-energy equation of state (EOS) from the cosmological-constant value are tested most effectively. Within a generalized CPL parameterization, we derive an analytic expression for the optimal redshift by maximizing the separation of the EOS from $w=-1$ relative to the corresponding uncertainty. We establish the optimal redshift relationship to the phantom-crossing and pivot redshifts. As an illustration, we apply the formalism to the dataset combination of DESI DR2 BAO measurements, the DES Year-6 independent BAO measurement, and the recalibrated DES-Dovekie supernova sample. Adopting the null hypothesis $\mathcal{H}_0:w(a_{\rm opt})=-1$, we find a tension of ~$2.8\sigma$ with the cosmological-constant prediction at the optimal redshift, compared with ~$2.6\sigma$ at the pivot redshift, despite the latter having a smaller EOS uncertainty. This behavior reflects the specific design of the optimal redshift to maximize the statistical significance of a departure of the EOS from the -1 value of the  $\Lambda$CDM model. We further provide statistical and geometrical interpretations of the optimum. In the CPL parameter space, the pivot corresponds to the projection that minimizes the variance of the equation of state, whereas the optimum maximizes its squared-distance from the cosmological-constant value over its variance. While the optimum is dataset and parameterization dependent, the underlying optimization principle and definition of the optimal redshift can be extended beyond the CPL framework. Future applications to DESI, Rubin LSST, Euclid, Roman Space Telescope, and other Stage-IV dark-energy surveys appear particularly promising. 
\end{abstract}

\maketitle


\section{Introduction}

The nature of dark energy remains one of the central open questions in modern cosmology, see e.g.  \cite{Carroll:2000fy,Peebles:2002gy,Copeland:2006wr,Ishak:2005xp}.  
Although the standard $\Lambda$ Cold Dark Matter ($\Lambda$CDM)  model has achieved remarkable success in describing a broad range of cosmological observations, ongoing efforts continue to focus on testing the assumption that the dark-energy component is a cosmological constant with an equation-of-state parameter fixed at $w=-1$. Current and forthcoming surveys aim not only to improve constraints on the present-day value of the dark-energy equation of state, but also to determine whether it evolves with cosmic time. In this context, the equation of state $w(z)$ has become a primary observable for characterizing possible departures from the cosmological-constant paradigm.

The question of whether dark energy may be dynamical and characterized by a time-dependent equation of state has received considerable attention after the Dark Energy Spectroscopic Instrument (DESI) collaboration reported tantalizing indications of departures from a cosmological constant. This came from DESI analyses of its first-year observations (Data Release 1, DR1) when combined with Cosmic Microwave Background (CMB), CMB lensing, and supernova (SN) data; see, for example, Refs.~\cite{DESI:2024mwx, DESI:2025zpo, ishak2024desi}. Subsequent DESI analyses based on the larger Data Release 2 (DR2) data set strengthened these indications and further increased interest in models with evolving dark-energy equations of state; see, e.g., Refs.~\cite{DESI:2025zgx,DESI:2025fii,DESI:2025gwf}. These developments triggered extensive activity in the literature, including reanalyses of the observational constraints, investigations of alternative dark-energy scenarios, studies of model dependence, and examinations of possible systematic effects. An incomplete list of such works is given in Refs.~\cite{Tada:2024xau, Wang:2024qan, Yin:2024xau, Luongo:2024xau, Cortes:2024xau, Colgain:2024xqj, Wang:2024rjd, Berghaus:2024kra, Wang:2024xau, Wang:2024pui, Shlivko:2024llw, Dinda:2024kjf, Bhattacharya:2024xau, Ramadan:2024xau, Roy:2024kni, Gialamas:2024lyw, Notari:2024xau, Liu:2024gfy, Orchard:2024xau, Hernandez-Almada:2024xau, Pourojaghi:2024xau, Giare:2024gpk, Jiang:2024xau, Efstathiou:2024xau, Reboucas:2024xau, RoyChoudhury:2024xau, Dhawan:2024xau, Linder:2024rdj, Park:2024xau2, Notari:2024xau2, Gao:2024ily, Fikri:2024klc, Tiwari:2024gzo, Tang:2024lmo, Zheng:2024xau, Odintsov:2024xau, Colgain:2024mtg, Lewis:2024xau, Koussour:2024xau, Sakr:2025fay, Yang:2025xau, Huang:2025xau, Wolf:2025xau, Giare:2025pzu, Sousa-Neto:2025gpj, Ormondroyd:2025exu, Khoury:2025txd, Ormondroyd:2025iaf, Brandenberger:2025hof, Nesseris:2025lke, Kessler:2025kju, You:2025uon, Shlivko:2025fgv, Cai:2025mas, Popovic:2025glk, Li:2025ops, Kou:2025yfr, Santos:2025wiv, DESI:2025wyn, Akarsu:2025dmj, Dinda:2025iaq, Wang:2025bkk, Mirpoorian:2025rfp, RoyChoudhury:2025dhe, Scherer:2025esj, Liu:2025mub, Chen:2025mlf, Chen:2025wwn, Efstathiou:2025tie, vanderWesthuizen:2025iam, Sabogal:2025jbo, Linder:2025zxb, Odintsov:2025jfq, Araya:2025rqz, Li:2025eqh, Herold:2025hkb, Lee:2025kbn, Chen:2025jnr, Hogas:2025ahb, Qiang:2025cxp, Lee:2025pzo, Chudaykin:2025aux, Silva:2025twg, Camarena:2025upt, Chaudhary:2025pcc, Wang:2025vtw, Yao:2025wlx, Arora:2025msq, Paul:2025wix, Toomey:2025xyo, Fazzari:2025lzd, Dinda:2025hiu, Park:2025fbl, RoyChoudhury:2025iis, Alam:2025epg, Artola:2025zzb, Reeves:2025xau, Yadav:2025vgo, Rezaei:2025vhb, Xu:2025nsn, Bolotin:2013jpa, Wang:2016lxa, Wang:2024vmw, Spec-S5:2025uom, Figueruelo:2026eis, Li:2025muv, Li:2024qso, Guedezounme:2025wav, Antusch:2026ldp, Bedroya:2025fwh, Andriot:2026lac, Anchordoqui:2026hys, Sailer:2025lxj, Jhaveri:2025neg, Elbers:2025xvk, Hergt:2026moc}.

A widely adopted framework for studying a time-varying equation of state is the CPL parameterization, which describes dark energy through the parameters $(w_0,w_a)$  ~\cite{Chevallier:2001,Linder2003}. These parameters respectively characterize the present-day value of the equation of state and its evolution with time. Because the two parameters are typically correlated, it is common to re-express the problem in terms of the pivot redshift, corresponding to the redshift at which a particular linear combination of $w_0$ and $w_a$ is measured with minimum uncertainty, see e.g.  \cite{Albrecht:2006um}. The pivot and its associated pivot equation of state has therefore become a standard quantity to explore in many dark-energy analyses.

In this paper-I, we introduce a new and complementary quantity that we refer to as the \emph{optimal redshift}. Rather than seeking the redshift at which the uncertainty on the equation of state is minimized, the objective of the \emph{optimum} is to identify the redshift at which the departure of the equation of state from a chosen reference model can be established with the greatest statistical significance. In the specific case considered here, the reference model is the cosmological-constant prediction $w=-1$. We derive an analytic expression for the optimal redshift, establish its relationship to the crossing and pivot redshifts, and show that it can be interpreted geometrically in the CPL parameter space as the redshift that maximizes the distance between the equation of state and the cosmological-constant value $w=-1$ when normalized by the corresponding  uncertainty. In a companion paper-II \cite{paper-II}, we apply the optimum framework to various current cosmological dataset combinations and investigate in detail its implications for the phantom-crossing behavior of the dark-energy equation of state.

The paper is organized as follows. In Sec.~II, we recall the definition of CPL parameterization and its generalization, and then derive the optimal redshift, together with its relation to the crossing and pivot redshifts. We also discuss several limiting cases. In Sec.~III, we illustrate the practical implementation of the formalism through a representative cosmological dataset, showing how the optimal redshift can be used to formulate and test a one-dimensional null hypothesis against the cosmological-constant prediction. We then provide visualization, geometrical interpretation, and a statistical interpretation of the optimum.  Finally, our conclusions and perspectives for future applications are presented in Sec.~IV.

\section{Dynamical dark energy and Optimal Redshift}

\subsection{Equations of state}

Dark energy can be modeled as a perfect fluid characterized by an energy density $\rho_{\rm DE}$ and a pressure $p_{\rm DE}$. Its equation of state (EOS) parameter is defined as $w(a)\equiv \frac{p_{\rm DE}}{\rho_{\rm DE} c^2}$, where $c$ is the speed of light (that we set equal to 1 to simplify notation) and $a$ is the cosmological scale factor, related to the redshift $z$ through $a=\frac{1}{1+z}$. The value and time evolution of the EOS provide a way to distinguish between different dark-energy models and possible departures from a cosmological constant. In particular, the cosmological constant corresponds to the special case, $w(a)=-1$. The time evolution of the dark-energy EOS is often described using the Chevallier-Polarski-Linder (CPL) parameterization ~\cite{Chevallier:2001,Linder2003}:  
\begin{equation}
w(a)=w_0+w_a(1-a),
\label{eq:cpl}
\end{equation}
where $w_0$ denotes the present-day value of the EOS (at $a=1$), while $w_a$ quantifies the amplitude of its time evolution, corresponding to minus the slope of $w(a)$ with respect to the scale factor. While the CPL parameterization is traditionally expressed about the present epoch ($a=1$), it can be generalized to an arbitrary scale factor $a_*$ that can be written as 
\begin{equation}
w(a)=w_*+w_a(a_*-a),
\label{eq:cplstar}
\end{equation}
where $w_*$ is the EOS evaluated at the reference scale factor $a_*$, $w_*\equiv w(a_*)$, and $w_a$ retains its interpretation as the parameter governing the time evolution of the EOS. We refer to Eq.~(\ref{eq:cplstar}) as CPL$_{*}$ parameterization, of which the standard CPL form is recovered by setting $a_*=1$.  Finally, using Eq.~(\ref{eq:cpl}) and Eq.~(\ref{eq:cplstar}), one finds the relation
\begin{equation}
w_* = w_0 + w_a(1-a_*).
\label{eq:wstarrelation}
\end{equation}
\subsection{Optimal redshift}
Since our goal is to identify the redshift at which the departure from the cosmological-constant value is most significant, we seek a point that simultaneously maximizes the distance of the EOS from $w=-1$ while minimizing the associated uncertainty. This can be formulated as the optimization problem. 
\begin{equation}
\max_{a_*} \; \mathcal{R}(a_*)
=
\max_{a_*} \;\frac{\left[w(a_*)+1\right]^2}
     {\langle \delta w_*^2\rangle},
\label{eq:Rdef}
\end{equation}
where $\langle \delta w_*^2\rangle$ denotes the variance of the EOS evaluated at the scale factor $a_*$ (throughout this work, angle brackets $\langle \cdot \rangle$ denote expectation values over the posterior distribution) and is given by 
\begin{equation}
\langle \delta w_*^2\rangle
=
\langle \delta w_0^2\rangle
+
(1-a_*)^2
\langle \delta w_a^2\rangle
+
2(1-a_*)
\langle \delta w_0 \delta w_a\rangle,
\label{eq:varwstar}
\end{equation}
where $\langle \delta w_0^2\rangle$ and
$\langle \delta w_a^2\rangle$ are the variances of
$w_0$ and $w_a$, respectively, and
$\langle \delta w_0 \delta w_a\rangle$
is their covariance.
Intuitively, we set \cref{eq:Rdef} such that the numerator measures the squared distance between the EOS and the cosmological-constant value $w=-1$, while the denominator quantifies the uncertainty associated with this measurement. Maximizing $\mathcal{R}$ therefore identifies the scale factor at which the departure from $\Lambda$CDM is most significant.  
Equivalently, this corresponds to maximizing the one-dimensional
$\Delta\chi^2$ between the EOS and the cosmological-constant value (see the right-hand side of \cref{eq:Rdef}).

The optimal scale factor is defined by
\begin{equation}
\left.
\frac{d\mathcal{R}}{da_*}
\right|_{a_*=a_{\rm opt}}
=0.
\label{eq:defopt}
\end{equation}
Substituting Eqs.~(\ref{eq:wstarrelation}) and (\ref{eq:varwstar}) into Eq.~(\ref{eq:defopt}) and solving for $a_*$ gives the optimal redshift
\begin{equation}
a_{\rm opt}
=
\frac{
w_a\big(\langle \delta w_0^2\rangle+\langle \delta w_0\delta w_a\rangle\big)
-(w_0+1)\big(\langle \delta w_a^2\rangle+\langle \delta w_0\delta w_a\rangle\big)
}
{
w_a\langle \delta w_0\delta w_a\rangle
-(w_0+1)\langle \delta w_a^2\rangle
}.
\label{eq:aopt}
\end{equation}
It is useful to examine how the optimal scale factor is related to two other characteristic scale factors that arise naturally within the CPL description. The first is the crossing scale factor, defined by the condition, $w(a_c)=-1$, which yields
\begin{equation}
a_c
=
1+\frac{w_0+1}{w_a},
\label{eq:ac}
\end{equation}
where $a_c$ denotes the scale factor at which the EOS crosses the phantom boundary. The second is the well-known pivot scale factor, obtained by minimizing the variance of the EOS:
\begin{equation}
\left.
\frac{d\langle \delta w_*^2\rangle}{da_*}
\right|_{a_*=a_p}
=
0,
\label{eq:pivotdef}
\end{equation}
which gives
\begin{equation}
a_p
=
1+
\frac{
\langle \delta w_0 \delta w_a\rangle
}
{
\langle \delta w_a^2\rangle
}.
\label{eq:ap}
\end{equation}
%

Using the definitions of the crossing scale factor, Eq.~(\ref{eq:ac}), and the pivot scale factor, Eq.~(\ref{eq:ap}), the expression for the optimal scale factor can be recast as
\begin{equation}
a_{\rm opt}
=
a_p
+
\frac{(a_p-1)^2
-
\frac{\langle \delta w_0^2\rangle}
       {\langle \delta w_a^2\rangle}}
     {a_c-a_p}.
\label{eq:aoptcompact}
\end{equation}
The corresponding equation-of-state value at the optimum can be written as
\begin{equation}
w_{\rm opt}=w(a_{\rm opt})
=
w_p
-
w_a
\frac{(a_p-1)^2
-
\frac{\langle \delta w_0^2\rangle}
       {\langle \delta w_a^2\rangle}}
     {a_c-a_p}. 
\label{eq:woptpivot}
\end{equation}
where $w_p=w(a_p)$ is the equation of state evaluated at the pivot scale factor.
Moreover, an equivalent form of \cref{eq:aoptcompact} can be written as 
\begin{equation}
a_{\rm opt} = a_p-\frac{{\rm det}(\mathbf{C})}{\langle \delta w_a^2\rangle^2(a_c-a_p)}
\label{eq:aoptwithC}
\end{equation}
where $\mathbf{C}$ denotes the $2\times2$ covariance matrix for $w_0$ and $w_a$.

Equations~(\ref{eq:aoptcompact}) and (\ref{eq:woptpivot}) make explicit the connection between the optimum, pivot, and crossing descriptions. While the pivot is constructed solely to minimize the uncertainty on the equation of state, the optimal point balances uncertainty minimization with maximal separation from the cosmological-constant value. Relative to the pivot, the optimum is shifted in a direction that increases the statistical significance of the departure from $w=-1$. Consequently, the optimum balances statistical precision against maximal sensitivity to deviations from the cosmological-constant prediction. The magnitude and sign of this shift are governed both by the separation between the pivot and crossing points and by the relative uncertainties associated with the amplitude and time evolution of the equation of state. Equations (\ref{eq:aoptcompact}) and (\ref{eq:woptpivot}) therefore show explicitly that the optimum is determined by a combination of geometry in parameter space and the underlying covariance structure of the dark-energy constraints.

As in the pivot formalism, once the parameters $w_0$ and $w_a$, together with their covariance matrix, have been estimated from observations, the quantities $a_{\rm opt}$ and $w_{\rm opt}$ can be computed directly without requiring any additional parameter estimation.

\subsection{Limiting Cases}
Additional insight into the nature of the optimum can be obtained by considering several limiting cases.
\\ \\ 
\textbf{a. No evolution:} $w_a \to 0$. In the limit of a constant equation of state, $w(a)\to w_0$, the crossing scale factor becomes ill-defined and the distinction between different projection directions disappears. In this regime, the optimum loses its significance since all scale factors yield the same equation of state.
\\ \\
\textbf{b. Vanishing covariance.} When $\langle\delta w_0\delta w_a\rangle = 0$, the pivot reduces to the present epoch, $a_p = 1$. In this limit, the covariance no longer contributes to the optimization, and the optimum is determined solely by the competition between the distance from the $w=-1$ boundary and the individual variances of $w_0$ and $w_a$.
\\ \\ 
\textbf{c. Pivot approaching the crossing.} For many current dataset combinations, the pivot and crossing scale factors are relatively close, $a_p\simeq a_c$. In this situation, the pivot equation of state is often close to the cosmological-constant value, $w_p\simeq -1$, limiting the statistical significance of the deviation despite the reduced uncertainty. Equation (11) shows that the shift between the optimum and pivot is inversely proportional to the separation $(a_c-a_p)$. Consequently, as the pivot approaches the crossing, the optimum becomes increasingly displaced from the pivot in order to increase the separation from $w=-1$, thereby keeping an enhanced overall statistical significance.
\\ \\ 
\textbf{d. Optimal and pivot coinciding.} Using \cref{eq:aoptwithC}, the difference between the optimum and pivot scale factors can be expressed as \[ a_{\rm opt}-a_p= -\frac{\det(C)} {\langle\delta w_a^2\rangle^2\,(a_c-a_p)}. \] For any physically realistic covariance matrix inferred from observations, the covariance matrix is positive definite and therefore has $\det(C)>0$. Consequently, the optimum and pivot are generically distinct and are almost never exactly equal. This can be understood directly from the above expression, which shows that a coincidence of the two can occur only in the limiting case of a singular covariance matrix. This is a consequence of the fact that the pivot minimizes the  variance of the EOS, whereas the optimum maximizes its statistical departure from the cosmological-constant value relative to the  uncertainty.
\subsection{Basis Dependence and Re-parameterizations}
The optimum redshift was derived within the CPL framework and therefore depends on the parameterization used to describe the dark-energy equation of state. This dependence is not unique to the optimum. Other commonly used quantities, including the pivot scale factor and pivot equation of state, are likewise parameterization dependent. Since both quantities are constructed from a specific choice of basis in parameter space, their numerical values generally change under a redefinition of the underlying parameters. Moreover, just like the pivot, the optimum is specific to a given dataset combination.  

The existence of an optimum, however, is more general than the CPL parameterization itself. The optimization criterion introduced in Eq.~(\ref{eq:Rdef}) can be applied to any parametrized equation of state for which both a mean prediction and an uncertainty can be assigned at a given redshift. In this broader context, the optimum corresponds to the redshift that maximizes the statistical separation from a chosen reference model, normalized by the uncertainty at that redshift.

For alternative dark-energy parameterizations, the function $w(z)$ and its covariance structure will differ from those of CPL. Consequently, the numerical value of $z_{\rm opt}$ may change. Nevertheless, the underlying principle remains unchanged: the optimum identifies the epoch at which a departure from the reference model can be established with the greatest statistical significance.

In the present work we focus on the CPL and CPL$_*$ parameterizations because they remain among the most widely used descriptions of dynamical dark energy and provide a convenient framework for deriving analytic expressions for both the pivot and optimal redshifts. However, this method can be applied to other parameterizations.  
%
\section{Illustrative Implementation of the Optimum Framework and its interpretations}
%
\subsection{Calculation of the Optimum and Tension Estimation Using a Dataset}

As an illustrative application of the framework developed in this work, we consider, for example,  the CPL constraints obtained in the  analysis \cite{Ishak:2025cay} using the combination of DESI DR2 BAO measurements, the DES Year-6 independent BAO measurement (No Overlap), together with the recalibrated DES-Dovekie Type-Ia supernova compilation. This dataset combination is denoted DESI+DESY6BAO (No Overlap)+DES-Dovekie SN as in Ref.~\cite{Ishak:2025cay}.

For this dataset combination, the CPL parameters were constrained to be
\begin{equation}
w_0=-0.840^{+0.065}_{-0.076},
\end{equation}
and
\begin{equation}
w_a=-0.52\pm0.49,
\end{equation}
with covariance matrix
\begin{equation}
\mathbf{C}
=
\begin{pmatrix}
0.004941 & -0.028877\\
-0.028877 & 0.236999
\end{pmatrix}.
\label{eq:cov}
\end{equation}

We can now derive the 
\begin{itemize}

\item Using Eq.~(\ref{eq:ap}), we obtain the pivot scale factor and corresponding pivot redshift:
\begin{equation}
a_p = 0.878,
\qquad
z_p = 0.139.
\end{equation}

\item Using Eq.~(\ref{eq:wstarrelation}), the corresponding pivot equation of state is
\begin{equation}
w_p = -0.903 \pm 0.038.
\end{equation}

\item Using Eq.~(\ref{eq:ac}), we find the phantom-crossing scale factor and corresponding redshift:
\begin{equation}
a_c = 0.692,
\qquad
z_c = 0.444.
\end{equation}

\item Using Eq.~(\ref{eq:aopt}) or Eq.~(\ref{eq:aoptcompact}), the optimal scale factor and corresponding redshift are
\begin{equation}
a_{\rm opt}=0.910,
\qquad
z_{\rm opt}=0.098.
\end{equation}

\item Using Eq.~(\ref{eq:woptpivot}), the equation of state evaluated at the optimum is
\begin{equation}
w_{\rm opt}
=
-0.887 \pm 0.041.
\end{equation}

\item For this dataset combination, the three characteristic redshifts satisfy
\begin{equation}
z_{\rm opt}<z_p<z_c,
\end{equation}
\begin{equation}
0.098 < 0.139 < 0.444.
\end{equation}
Thus, all three points lie within the observationally accessible redshift range, with the optimum occurring at a slightly lower redshift than the pivot point.

\end{itemize}

We now formulate the null hypothesis
\begin{equation}
\mathcal{H}_0:\qquad w_{\rm opt}=-1,
\end{equation}
which corresponds to the cosmological-constant prediction for the EOS when evaluated at the optimal scale factor. The significance of the departure from this null hypothesis is computed using Eq.~(\ref{eq:Rdef}),
\begin{equation}
\mathcal{R}(a_{\rm opt}) = - \Delta\chi^2_{\rm opt} = 
\frac{\left(w_{\rm opt}+1\right)^2}
{\langle \delta w_{\rm opt}^2\rangle}
=
7.71,
\end{equation}
which corresponds to\footnote{We follow the same convention as in \cite{DESI:2025zgx} where a negative $\Delta\chi^2$ signifies the preference of the alternative model over the $\Lambda$CDM model.}
\begin{equation}
n_\sigma
=
\sqrt{|\Delta\chi^2_{\rm opt}|}
=
2.78 .
\end{equation}

For comparison, applying the same procedure at the pivot point gives
\begin{equation}
\mathcal{R}(a_p) = - \Delta\chi^2_p
=
\frac{\left(w_p+1\right)^2}
{\langle \delta w_p^2\rangle}
=
6.57,
\end{equation}
corresponding to
\begin{equation}
n_{\sigma,p}=2.56 .
\end{equation}

Therefore, for this dataset combination, the equation of state evaluated at the optimal scale factor differs from the cosmological-constant prediction by approximately $2.8\sigma$, compared to approximately $2.6\sigma$ at the pivot point. This result illustrates the key distinction between the pivot and optimum constructions. By definition, the pivot scale factor minimizes the projected variance of the equation of state and therefore provides the smallest uncertainty. In the present example,
\begin{equation}
\sigma(w_p)=0.038
<
\sigma(w_{\rm opt})=0.041.
\end{equation}

Nevertheless, the optimum yields a larger significance. This occurs because the optimization criterion is not designed to minimize the uncertainty alone, but rather to maximize the ratio between the distance from the cosmological-constant value and the corresponding uncertainty. In other words, the shift of the mean value away from $w=-1$ when moving from the pivot to the optimum more than compensates for the modest increase in the  uncertainty. As a result,
\begin{equation}
n_{\sigma,\rm opt}
=
2.78
>
2.56
=
n_{\sigma,p},
\end{equation}
demonstrating explicitly that the optimum identifies the redshift at which the departure from the cosmological-constant prediction is statistically most significant.
This significance corresponds to a one-dimensional test of the null hypothesis $w(a_{\rm opt})=-1$ and follows directly from the optimal-redshift equations. 
This example is intended only to illustrate the practical implementation of the optimum framework. A more comprehensive study using a variety of dark-energy constraints and dataset combinations is presented in a companion paper \cite{paper-II}. 
\begin{figure}[t]
\centering
\includegraphics[width=0.60\columnwidth]{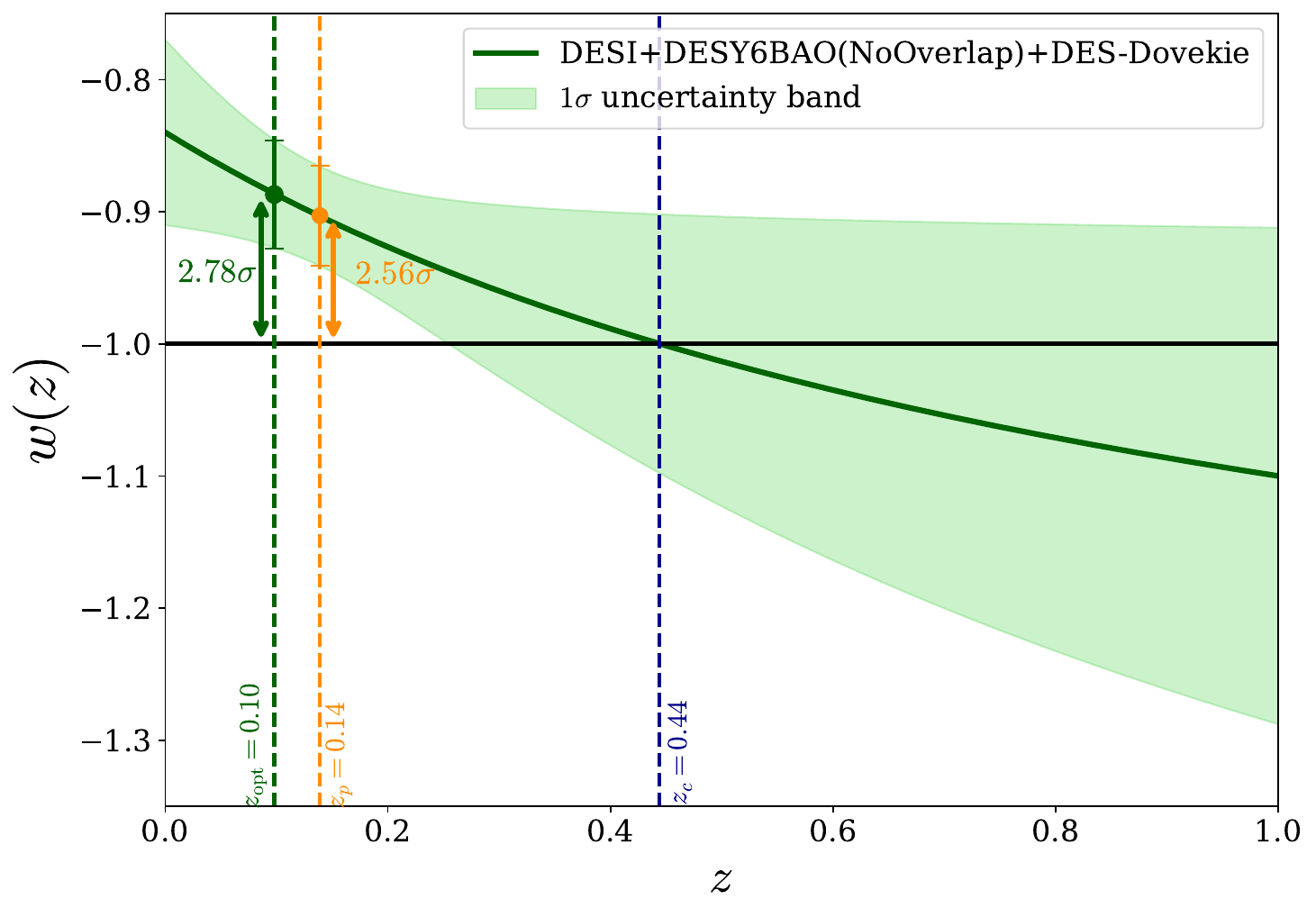}
\caption{
Illustration of the Optimum redshift using the evolution of the CPL equation of state $w(z)$ for the DESI+DESY6BAO (No Overlap)+DES-Dovekie SN dataset combination. The solid dark-green curve shows the mean equation of state, while the light-green shaded band represents the propagated $1\sigma$ uncertainty obtained from the covariance matrix of $(w_0,w_a)$. The horizontal black line indicates the cosmological-constant value $w=-1$. The vertical dashed lines mark the phantom-crossing redshift $z_c=0.44$ (dark blue), the pivot redshift $z_p=0.14$ (orange), and the optimal redshift $z_{\rm opt}=0.10$ (dark green). The orange and dark-green markers indicate the equation-of-state values and associated uncertainties at the pivot and optimum points, respectively. Although the pivot point minimizes the projected uncertainty, it lies closer to the cosmological-constant boundary and therefore corresponds to a significance of only $2.56\sigma$. By contrast, the optimal point occurs where the ratio between the displacement from $w=-1$ and the   uncertainty is maximized. The corresponding $w_{\rm opt}$ yields a larger significance of $2.78\sigma$, despite having a slightly larger uncertainty. The vertical double arrows illustrate the displacements $|w_p+1|$ and $|w_{\rm opt}+1|$ that enter the one-dimensional tension estimator ratio $\mathcal{R}(a)$ in \cref{eq:Rdef} that determine the statistical significance of the departure from the cosmological-constant prediction.
}
\label{fig:woptexample}
\end{figure}
%

\subsection{Visualization and Geometrical Interpretation}

The origin of the improvement in the tension at the optimal redshift can be understood both from the evolution of the equation of state and from the geometry of the CPL parameter space. Figure \ref{fig:woptexample} shows the evolution of the equation of state together with its uncertainty band for the dataset combination considered in the previous subsection. The crossing, pivot, and optimal redshifts are indicated on the figure. Although the pivot point provides the smallest  uncertainty on the equation of state, the figure shows that it does not necessarily correspond to the largest statistical separation from the cosmological-constant value $w=-1$. In the present example, the pivot value remains relatively close to the $w=-1$ boundary despite its smaller uncertainty. By contrast, the optimal point occurs at a nearby redshift where the equation of state has evolved further away from the cosmological-constant value. The increase in this separation then exceeds the compensation for the modest increase in the uncertainty, resulting in a larger significance level.
This behavior is quantified by the optimization criterion introduced in Eq.~\ref{eq:Rdef}, which measures the squared distance from the cosmological-constant prediction in units of the corresponding variance. This was analytically expressed using the quantity $\mathcal{R}(a)$. The optimal scale factor therefore identifies the epoch at which the equation of state is most significantly separated from the cosmological-constant prediction.

\begin{figure}[t]
\centering
\includegraphics[width=0.60\columnwidth]{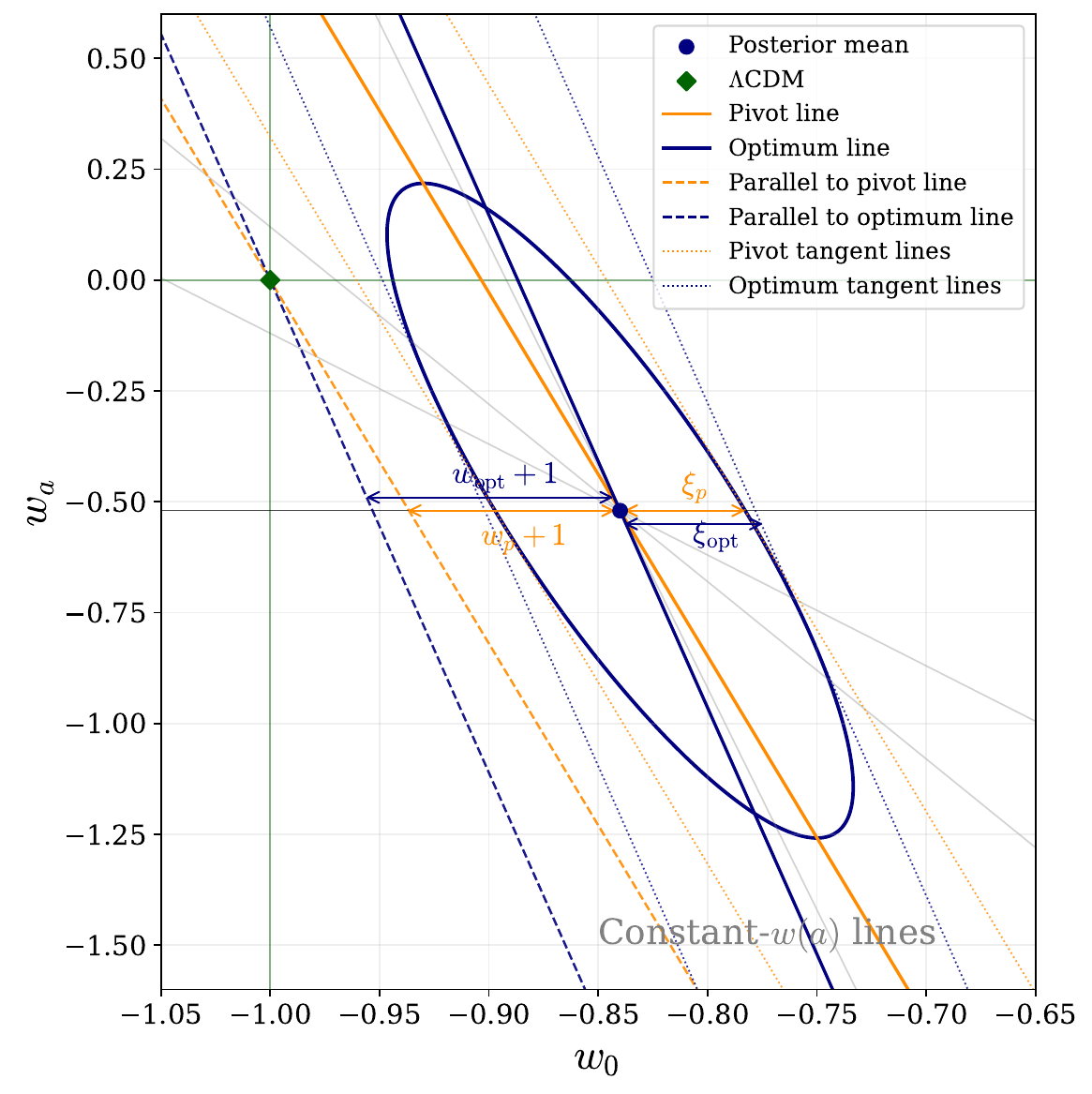}
\caption{ Geometrical interpretation of the pivot and optimum projections in the CPL parameter space. The black ellipse is the 68\% confidence region (\(\Delta\chi^2=2.30\)) derived from the covariance matrix. The solid orange and blue lines denote the constant-\(w_p\) and constant-\(w_{\rm opt}\) contours passing through the posterior mean, while the dotted parallel lines are the corresponding tangent lines to the ellipse. Along the horizontal line \(w_a=\bar w_a\), the segments \(\xi_p\) and \(\xi_{\rm opt}\) are proportional to the projected uncertainties \(\sigma(w_p)\) and \(\sigma(w_{\rm opt})\), respectively. The segments \(w_p+1\) and \(w_{\rm opt}+1\) measure the displacement of the pivot and optimum contours from the cosmological-constant contour (\(w=-1\)). Although \(\xi_{\rm opt}\) is slightly larger than \(\xi_p\), indicating a modest increase in projected uncertainty, the substantially larger value of \(|w_{\rm opt}+1|\) compared to \(|w_p+1|\) more than compensates for this increase. Consequently, the optimum yields a larger value of \(|w(a)+1|/\sigma(w(a))\) and therefore a more statistically significant departure from the cosmological-constant prediction. }
\label{fig:geometry}
\end{figure}

Additional insight is provided by the geometrical representation shown in Fig.~\ref{fig:geometry}. In the CPL parameter space, every scale factor defines a particular linear combination of \(w_0\) and \(w_a\). Equivalently, each scale factor corresponds to a family of straight lines, 
\begin{equation} 
w_0+(1-a)w_a = {\rm constant}, 
\label{eq:constantwlines} 
\end{equation} 
whose orientation varies continuously with \(a\). The covariance ellipse shown in Fig.~\ref{fig:geometry} represents the posterior distribution of the CPL parameters obtained from the dataset considered above. Different choices of the scale factor correspond to different projections of this covariance ellipse. The projected width of the ellipse determines the uncertainty on \(w(a)\), while the projected displacement from the cosmological-constant boundary contributes to the determination of the statistical significance of the departure from \(w=-1\). The uncertainty on \(w(a)\) is not associated with the constant-\(w(a)\) line itself, but rather with a direction that crosses it. Consequently, each choice of scale factor corresponds to a different projection of the covariance ellipse and therefore to a different uncertainty \(\sigma(w(a))\). 

It is important to note that the transformation from the basis \((w_0,w_a)\) to \((w_p,w_a)\), where \(w_p\equiv w(a_p)\), is not an orthogonal rotation but a shear transformation. As a result, although the variances of \(w_p\) and \(w_a\) are represented directly by the semi-axes of the ellipse in the \((w_p,w_a)\) basis, the same is not true in the \((w_0,w_a)\) plane. A similar consideration applies to \(w_{\rm opt}\). We therefore introduce geometrical constructions that allow us to compare quantities proportional to \(\sigma(w_p)\), \(\sigma(w_{\rm opt})\), (\(w_p+1\)), and (\(w_{\rm opt}+1\)). 

As shown in Fig.~\ref{fig:geometry}, we first plot the 68\% confidence ellipse (\(\Delta\chi^2=2.30\)) associated with the covariance matrix. We then plot the constant-\(w\) lines corresponding to the pivot and optimum in solid orange and solid blue, respectively, passing through the posterior mean. Note that the pivot line is not generally aligned with the semi-major axis because of the shear transformation discussed above. Using the same colors, we then draw the corresponding tangent lines parallel to these constant-\(w\) directions. To connect the projected uncertainties to geometrical quantities, we define
\begin{equation} 
\boldsymbol{\theta} = \begin{pmatrix} w_0-\bar w_0\\[2mm] w_a-\bar w_a \end{pmatrix}, \qquad \mathbf{v} = \begin{pmatrix} 1\\[2mm] 1-a \end{pmatrix}. \end{equation} 
The family of constant-\(w(a)\) lines may then be written as 
\begin{equation} 
\mathbf{v}^{T}\boldsymbol{\theta}=q, \end{equation} 
or equivalently 
\begin{equation} 
(w_0-\bar w_0)+(1-a)(w_a-\bar w_a)=q. 
\end{equation} 
For a fixed value of \(a\), the parameter \(q\) labels a family of parallel constant-\(w(a)\) lines. One can show, either by imposing the tangency condition between the ellipse and the corresponding constant-\(w(a)\) line, or equivalently by means of a Lagrange-multiplier calculation, that the tangent member of this family satisfies 
\begin{equation} 
q_{\rm tangent} = \sqrt{\Delta\chi^2}\, \sigma(w(a)),
\end{equation} 
hence, for the 68\% confidence ellipse, 
\begin{equation} 
q_{\rm tangent} = \sqrt{2.30}\, \sigma(w(a)). 
\label{eq:qsigma} 
\end{equation} 
The next step is to identify a geometrical measure directly proportional to the  uncertainty. Evaluating Eq.~(\ref{eq:qsigma}) on the horizontal line passing through the posterior mean point 
\begin{equation} 
w_a=\bar w_a, 
\end{equation} 
gives 
\begin{equation} 
\xi = w_0-\bar w_0 = \sqrt{2.30}\, \sigma(w(a)). 
\end{equation} 
Although one may be tempted to use the perpendicular distance from the center of the ellipse to the tangent line, that quantity contains a dependence on the the scale factor and therefore is not proportional solely to \(\sigma(w(a))\). 
We therefore adopt \(\xi\), measured along the horizontal line \(w_a=\bar w_a\), as a geometrical estimator proportional to the  uncertainty where the dependence on a is removed.  

As seen in Fig.~\ref{fig:geometry}, 
\begin{equation} 
\xi_p < \xi_{\rm opt}, 
\end{equation} 
reflecting 
\begin{equation} 
\sigma(w_p) < \sigma(w_{opt}), 
\end{equation} 
and the defining property of the pivot as the minimum-variance projection. 

Next we consider the separation between the constant-\(w\) line passing through the posterior mean and the parallel line passing through the cosmological-constant point \((-1,0)\). For a given scale factor \(a_*\), these lines are 
\begin{equation} 
w_0+(1-a_*)w_a=w_*, 
\end{equation} and 
\begin{equation} 
w_0+(1-a_*)w_a=-1. 
\end{equation} 
Evaluating the separation between these two parallel lines along the same horizontal line \(w_a=\bar w_a\) yields directly \begin{equation} 
|w_*+1|. 
\end{equation} 
Therefore, the geometrical quantities labeled \(w_p+1\) and \(w_{\rm opt}+1\) in Fig.~\ref{fig:geometry} provide direct measures of the displacement from the cosmological-constant line to the pivot and optimum projection lines, respectively.

Consequently, the optimum is characterized by a slightly larger  uncertainty than the pivot, 
\begin{equation} 
\sigma(w_{\rm opt}) > \sigma(w_p), 
\end{equation} 
but also by a larger displacement from the cosmological-constant contour, 
\begin{equation} 
|w_{\rm opt}+1| > |w_p+1|. 
\end{equation} 
In view of the optimization criterion introduced in Eq.~(\ref{eq:Rdef}), the increase in separation outweighs the increase in uncertainty, thereby producing a larger value of \(\mathcal R(a)\) and a larger statistical significance at the optimum than at the pivot.

The geometrical interpretation shows that the optimum corresponds to the projection that maximizes the statistical separation from the cosmological-constant contour. Although it is associated with a slightly larger  uncertainty than the minimum-variance projection, it is characterized by a sufficiently larger displacement from the \(w=-1\) contour that the ratio \(|w(a)+1|/\sigma(w(a))\) is maximized. Consequently, the optimum identifies the redshift at which departures from the cosmological-constant prediction can be established with the greatest statistical significance.


\section{Conclusions}

In this work, we introduced and developed the concept of the optimal redshift. Within a generalized CPL parameterization centered on an arbitrary scale factor, we identified the redshift at which the equation of state exhibits the largest statistically significant departure from the cosmological-constant value. The resulting quantity emerges naturally from an optimization problem that balances the displacement from $w=-1$ against the associated uncertainty, thereby providing a statistically motivated redshift at which departures from $\Lambda$CDM can be tested most effectively. 

We also established explicit relationships between the optimal redshift, the pivot redshift, and the phantom-crossing redshift within the generalized CPL framework. These characteristic redshifts emerge from different physical or statistical criteria, yet are connected through simple analytical expressions.  The formalism further relates the corresponding equation-of-state values evaluated at these points. 
We analyze how these points evolve in various limiting cases. In particular, we find that for any physically realistic positive-definite covariance matrix in the $(w_0,w_a)$ parameter space, the optimum and pivot redshifts are generically distinct. This follows directly from the fact that the pivot minimizes the  variance of the EOS, whereas the optimum maximizes its statistical separation from the cosmological-constant value relative to the  uncertainty.

To illustrate the practical use of the optimal redshift formalism, we considered a representative cosmological dataset combination consisting of DESI DR2 BAO measurements, the DES Year-6 independent BAO measurement, and the recalibrated DES-Dovekie supernova sample. 
The optimal-redshift can be computed once the CPL parameters and covariance matrix are known like, for example, from the usual cosmological inference. We then formulated a one-dimensional null hypothesis: $\mathcal{H}_0:\quad w(a_{\rm opt})=-1$, corresponding to the cosmological-constant prediction evaluated at the optimal redshift. Applying the formalism to this dataset, we found a tension of approximately $2.8\sigma$ with the cosmological-constant value at the optimum redshift, compared with approximately $2.6\sigma$ at the pivot redshift. This explicitly demonstrates that minimizing the uncertainty alone does not necessarily maximize the significance of a departure from $\Lambda$CDM. Rather, the optimal redshift identifies the epoch at which the balance between displacement from $w=-1$ and statistical uncertainty produces the largest significance.

We further provided geometrical and statistical interpretations of the optimum. In the CPL parameter space, the pivot corresponds to the projection of the covariance ellipse that minimizes the  variance of the equation of state, whereas the optimum corresponds to the projection that maximizes its statistical separation from the cosmological-constant boundary relative to the  uncertainty, thereby maximizing the significance of the departure from $w=-1$.  In situations where the pivot point occurs close to a crossing point or other reference boundary, the optimum may provide a substantially more sensitive diagnostic than the pivot itself.

The framework also has limitations. Like the pivot redshift, the optimum redshift is dataset dependent and generally parameterization dependent, since it is derived from a specific representation of the dark-energy equation of state and its covariance structure. Although the optimization principle itself is more general, the numerical values of $a_{\rm opt}$ and $w_{\rm opt}$ will vary across different datasets and alternative dark-energy parameterizations. 

Looking ahead, the optimal-redshift framework should be useful as future surveys deliver tighter constraints on dark energy and its evolution. In particular, it may provide a valuable complement to conventional parameter estimation by identifying the redshifts at which competing cosmological models can be most effectively distinguished. Applications to future DESI analyses and other  Stage-IV cosmological surveys such as Rubin LSST, Roman Space Telescope, Euclid, and others appear promising.

Finally, the present paper-I focused on the formalism, interpretation, and a practical implementation of the optimal-redshift framework itself. In companion paper-II \cite{paper-II}, we apply the framework to a broad range of current cosmological dataset combinations, with particular emphasis on the phantom-crossing behavior of the dark-energy equation of state. 

\section{Acknowledgments}
We thank David Shlivko and Kushal Lodha for providing useful comments on the manuscript. The authors acknowledge that Copilot was used to proofread and improve the text in this paper, as well as generating scripts for the two plots in this paper. MI acknowledges that this material is based upon work supported in part by the Department of Energy, Office of Science, under Award Number DE-SC0022184 and also in part by the U.S. National Science Foundation under grant AST2327245.


\bibliography{refs_key_paper,DESI2024,DDEComment, Leo,INSPIRE-CiteAll_filtered_sorted, Refrences}

\end{document}